\RequirePackage{lineno}
\documentclass[preprint,12pt,3p,hyperref]{elsarticle}

\usepackage{graphicx}
\usepackage{epsfig}
\usepackage{amssymb}
\usepackage{geometry}
\usepackage{enumitem}
\usepackage{natbib}
\usepackage{subcaption}

\usepackage[colorlinks,citecolor=blue]{hyperref}

\journal{Chin. J. Space Sci}
\begin{document}
\begin{frontmatter}
\title{A cosmic microscope to probe the Universe from Present to Cosmic Dawn - dual-element low-frequency space VLBI observatory \footnote{The work is supported with funding from the Ministry of Science and Technology of China (grant No.2018YFA0404600) and the Chinese Academy of Sciences (CAS, grant No. 114231KYSB20170003). }}

\author[label1]{Tao An\corref{cor1}}
\cortext[cor1]{Corresponding author}
\ead{antao@shao.ac.cn} 

\author[label1]{Sumit Jaiswal}
\author[label1]{Prashanth Mohan}
\author[label1]{Zhen Zhao}
\author[label1]{Baoqiang Lao}

\address[label1]{Shanghai Astronomical Observatory, Chinese Academy of Sciences, \\80 Nandan Road, Shanghai 200030, China}


\begin{abstract}
A space-based very long baseline interferometry (VLBI) programme, named as the Cosmic Microscope, is proposed to involve dual VLBI telescopes in the space working together with giant ground-based telescopes (e.g., Square Kilometre Array, FAST, Arecibo) to image the low radio frequency Universe with the purpose of  unraveling the  compact structure of cosmic  constituents including supermassive black holes and binaries, pulsars, astronomical masers  and the underlying source, and exoplanets  amongst others. The operational frequency bands are 30, 74, 330 and 1670 MHz, supporting broad science areas. The mission plans to launch two 30-m-diameter radio telescopes into 2,000 km $\times$ 90,000 km elliptical orbits. The two telescopes can work in flexibly diverse modes: (i) space-ground VLBI. The maximum space-ground baseline length is about 100,000 km; it provides a high-dynamic-range imaging capacity with unprecedented high resolutions at low frequencies (0.3 mas at 1.67 GHz and 20 mas at 30 MHz) enabling studies of exoplanets and supermassive black hole binaries (which emit nanoHz gravitational waves); (ii) space-space single-baseline VLBI. This unique baseline enables the detection of flaring hydroxyl masers, and more precise position measurement of pulsars and radio transients at mas level; (iii) single dish mode, where each telescope can be used to monitor transient bursts and rapidly trigger follow-up VLBI observations. The large space telescope will also contribute in measuring and constraining the total angular power spectrum from the Epoch of Reionization. In short, the Cosmic Microscope offers astronomers the opportunity to conduct novel, frontier science. 

\end{abstract}


\end{frontmatter}

\section{Background} 
\label{sec:background}

The Very Long Baseline Interferometry (VLBI) technique was initially employed to get high resolution images and unravel the fine structure of compact quasars \cite{ref1}. The angular resolution of any radio interferometer is determined by the observing wavelength and the length of the baseline (the projected separation of two elements of the interferometer). The observing waveband is set by the science goal and the receiver/instrumental capability. The angular resolution of ground-based VLBI is limited by the size of the Earth (around 10,000 km); this may not be good enough to study many compact sources. To vastly improve the angular resolution by increasing the baseline length, the space VLBI technique was devised \cite{ref2,ref3,ref4,ref5,ref6,ref7,ref8,ref9}.

The space VLBI uses at least one space-based radio telescope along with ground-based radio telescopes to form a space-ground VLBI network, or two space-based radio telescopes to enhance the space-ground ($u,v$) coverage, or three and more space-based telescopes constituting a space-space VLBI network. Space VLBI observations can enable the imaging of radio sources with unprecedented angular resolutions (a few times better than that with the ground-based VLBI) which far exceeds (by orders of magnitude) other astronomical measurements. 

The first dedicated space VLBI mission was the VLBI Space Observatory Program (VSOP\cite{ref10}) led by Japan and launched in 1997 February.  It was successfully operated for more than 6 years and consisted of an 8 m diameter radio telescope positioned in an elliptical Earth orbit with an apogee height of 21,400 km, forming a maximum baseline three times the size of the Earth. VSOP is operational at 1.6 and 5 GHz. Due to the small space antenna size, the sensitivity of VSOP was however limited to observe only bright radio sources such as active galactic nuclei (AGN), pulsars and hydroxyl (OH) masers \cite{ref11}.  

In 2011 July, the Spektr-R (RadioAstron) mission \cite{ref12} was launched by the Russian Space Agency. It consists of a 10 m diameter space radio telescope which, with a maximum baseline length of 340,000 km offers a high angular resolution of about eight microarcsecond \cite{ref13}, thus setting a milestone in the development of space VLBI. The observing frequency bands of the RadioAstron are 330 MHz, 1.67 GHz, 4.85 GHz and 22 GHz, employed mainly in the study of quasars, pulsars and cosmic masers. Other scientific objectives include determining the structure of interplanetary and interstellar plasma and studying the Earth's gravitational field.

A few other space VLBI missions are in development or proposed \cite{ref14,ref15,ref16} by the different institutions around the world. Most of these missions plan to observe at higher frequencies (up to 86 GHz) to realize higher angular resolutions. However, there are many astrophysical sources and phenomena exclusive to low radio frequencies \cite{ref17} such as the magnetospheric radio emission from Jupiter-like exoplanets. Such observations are however limited by absorption or refraction of the ionosphere for ground VLBI stations thus necessitating space-based observations. A low-orbit space VLBI mission RAKSAS with a 30-m orbital radio telescope was once proposed \cite{ref18}. Low radio frequency space VLBI in particular can provide unique contributions to supplement existing high-resolution radio astronomy. 

A low-frequency space VLBI mission with dual space telescopes is thus proposed here. It can be operated in three different modes, serving broad science cases. Section~\ref{sec:concept_tech} presents the overall concept design of the mission and key techniques. Section~\ref{sec:science_objectives} introduces the scientific objectives.  A summary is given in Section~\ref{sec:summary }. 

\section{Mission concept and key techniques}
\label{sec:concept_tech}

\begin{figure}[!t]
    \centering
    \includegraphics[width=0.8\columnwidth]{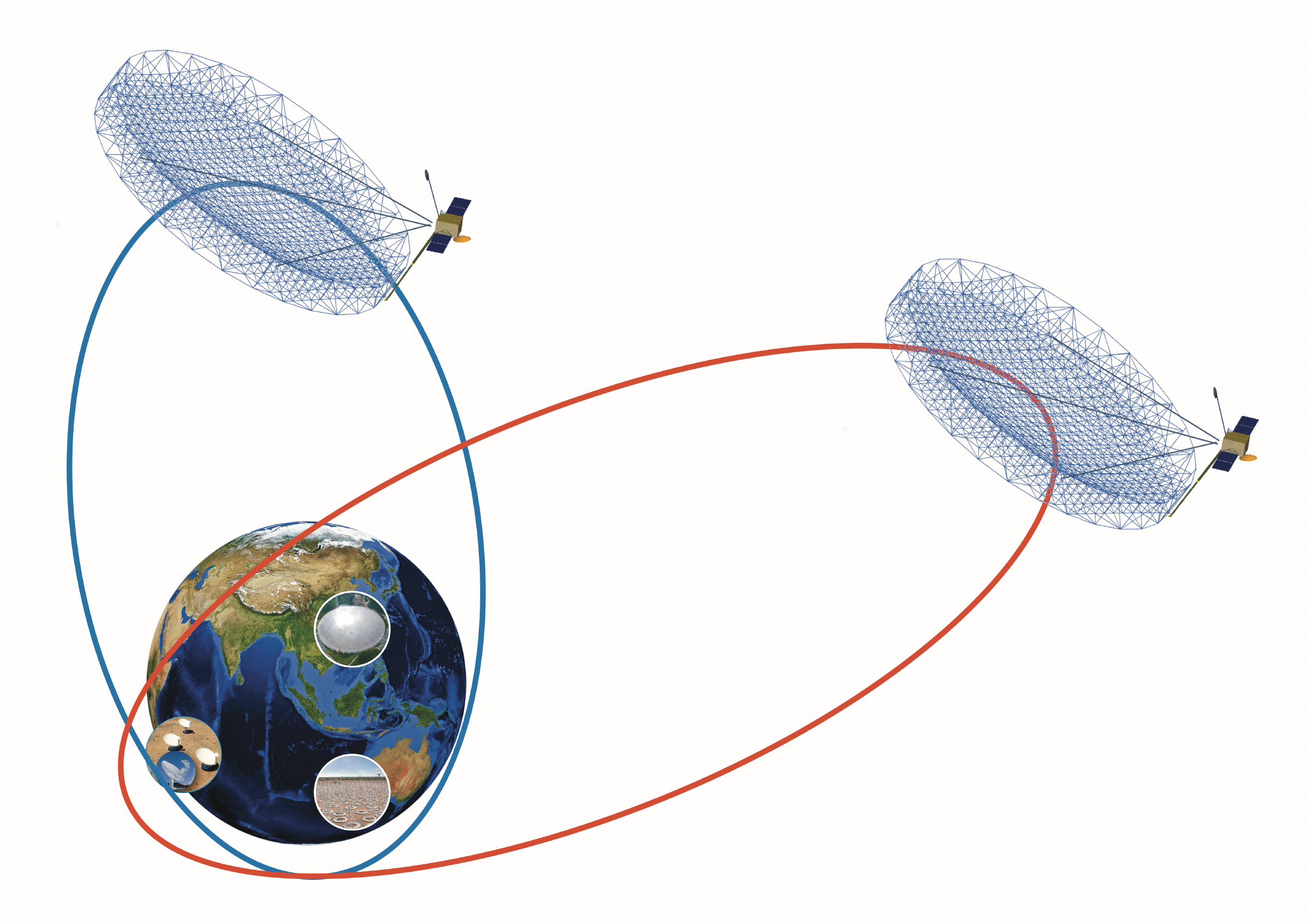}
     \caption{The concept of the Cosmic Microscope Mission. It involves two space-based telescopes with diameter of 30 m or even larger, placed in elliptical orbits, with apogee height of 90,000 km and 60,000 km, respectively. The space telescopes can complement a ground-based low-frequency VLBI network in which the FAST 500m, SKA1-Mid and SKA1-Low play an important role. }
    \label{fig:cmm}
\end{figure}

\textit{Cosmic Microscope} (CM) is a space VLBI mission concept proposed by the Shanghai Astronomical Observatory of the Chinese Academy of Sciences. It comprises two space-based radio telescopes with diameter of 30 m. The equivalent illumination area of the off-axis space telescope is similar with an elliptical shape of 30 m $\times$ 27 m. The two satellites are positioned in large elliptical orbits, with apogee heights of 90,000 km and 60,000 km, respectively and a perigee height of 2,000 km. The orbit inclination angles are roughly perpendicular when the second satellite is launched. Figure~\ref{fig:cmm} depicts a sketch of the proposed mission. 

Compared to previous space VLBI missions, the CM is uniquely operational at low radio frequencies between 30 MHz and 1.7 GHz. The lowest frequency band of VSOP is 1.67 GHz and that of RadioAstron is 330 MHz. 
The low frequency (below 330 MHz) radio regime has been largely unexplored with high resolution at mas resolutions.  

Ground radio interferometers, including the Very Large Array (VLA), LOW-Frequency ARray (LOFAR) and Murchison Widefield Array (MWA), have made low-frequency surveys at 10--240 MHz with resolutions of tens of arcsec\citep{VLA74MHz,LOFARsurvey,MWAsurvey}. 
However the Earth's ionosphere strongly attenuates extra-terrestrial radio signals at frequencies below 100 MHz (wavelength longer than 3 m), and terrestrial observatories are often affected by radio frequency interference, motivating the ultra long wavelength observatories to move into space. 
There have been some conceptual and prototype studies of space-based radio interferometers comprising an array of microsatellites each deployed with three orthogonal dipole antennas working at ultra long wavelengths (1--30 MHz) \citep[][and references therein]{DSL, Jia2018, An2019}. But the maximum baselines of these proposed interferometers are 30--100 km, resulting in relatively lower resolutions of arcminutes to degrees. 
The proposed VLBI capacity of the CM enables observations at ultra long wavelengths with mas resolutions, potentially producing novel scientific discoveries requiring high resolution.

In addition, the CM may gain strong support from the large low-frequency radio telescopes and arrays on the Earth which started commissioning recently or are being built.
Contemporary large radio telescopes will include the Five-hundred-meter Aperture Spherical radio Telescope (FAST), whose construction was finished in 2016, and the Square Kilometre Array that is expected to start construction in early 2020. Within the next 5-10 years, FAST and the first phase of SKA (SKA1) will be fully operational, providing strong synergy and support to the space VLBI telescope. This is thus an opportune epoch to consider the design of a low frequency space VLBI mission.

The large space-based baselines offer high angular resolution; sensitivity is however inhibited by poor Fourier plane sampling.   It can be improved by increasing the number of antennas, their sizes, observation bandwidth, and integration time. The CM mission is therefore planned to include two space-based radio telescopes with many advantages over those previously operated and/or proposed ones. First, two space telescopes increase the ($u,v$)  coverage on the sparse space-ground baselines. This is critical for improving image quality. Second, each space telescope will be 30 m in size, 3 times larger than RadioAstron.  As they can be deployed in synergy with the largest ground-based radio telescopes, such as FAST, Arecibo, SKA1-Mid, SKA1-Low, Effelsberg, GBT, GMRT, JVLA, and QTT, the space-ground VLBI network will possess an unmatched high sensitivity and resolution. Third, CM contains two space radio telescopes and will provide the advantage of multiple operational modes: using them as the space-ground VLBI or as the space-space VLBI or as the two-single space-based radio telescopes. The preferred mode of operation out of these three will be decided by the science goal. The proposed space VLBI is an unique concept of its kind, as the other space VLBI missions include only one space radio telescope. 

As we found that almost all those previous and contemporaneous space VLBI missions are generally focused on the high radio frequency regime, it is imperative for a mission that can cover the low frequency regime, enabling both complementary and novel science objectives. The proposed space VLBI is therefore planned to operate in the low frequency bands centered at 30, 74, 330 MHz and 1.67 GHz. The 30 and 74 MHz are mainly for Epoch of Reionization (EoR) observations and detecting auroral radiation from gas rich exoplanets. The 330 MHz can be used for observations and localization of pulsars, Supernova remnants, transients (including neutron star merger, fast radio bursts, gamma-ray bursts) and their afterglows. The highest frequency 1670 MHz is for the hydroxyl (OH) maser line observation; it can also be used for supermassive black hole binaries (SMBHB) search, and accurate astrometry. 

\begin{table}[!t]
\centering
\caption{parameters of satellites}\label{tab:tab1}
\begin{tabular}{|c|c|}
\hline
Weight  & 1500-2000 kg \\\hline
Payload weight & 800 kg \\\hline
Telescope & 30 m $\times$ 2 \\\hline
Power consumption & 2 kW \\\hline
Envelope & 3.6 m $\times$ 6.0 m \\\hline
Expected Lifetime & 8 years \\\hline
Propellant  & 150 kg \\\hline
Orbit 1 (CM1) & 90000 $\times$ 2000 km, 28.5$^\circ$ \\\hline
Orbit 2 (CM2) & 60000 $\times$ 2000 km, 148.5$^\circ$ \\\hline
 & Geostationary Transfer Orbit \\\hline
Rocket & Long Match (CZ-3C)\\
\hline
\end{tabular}
\end{table}

The parameters of the satellite platform are listed in Table~\ref{tab:tab1}. The payloads include space-based telescope, receiver, digital data acquisition and recording, time and frequency, and downlink systems. 

The key technical requirements of the proposed space VLBI mission include:
\begin{enumerate}
\item[1)] Unlike previous space VLBI, the proposed CM mission includes two space-based antennas of 30-m diameter each (high sensitivity, synergy with largest ground-based telescopes). Their manufacture is a major challenge. Moreover, the telescopes require the satellite platform to have rapid altitude maneuver and stabilization capabilities. 

\item[2)] One of the major challenges for low-frequency radio observations is the strong radio frequency interference (RFI) which degrade the data quality and even can cause data loss. For the CM mission, besides terrestrial RFI, that generated by the onboard electronics can also cause signal contamination. Electromagnetic compatibility (EMC) and shielding thus play a critical role.

\item[3)] This is a concept proposal for a space VLBI mission. With the advancement of technology, the launch of 30 m radio telescope will be achievable in near future.  A coordination VLBI network of CM and giant ground telescopes is highly desirable to obtain better quality images for detecting the faint emitting structures in target sources.

\item[4)] Both telescopes must observe the target source simultaneously to achieve an interferometric output. This requires stable and precise maneuvering time keeping, especially in case of space-ground VLBI. The mission needs the highly stable time/frequency standard and fast data transfer to the ground stations.

\item[5)] The space radio telescope sensitivity at 1.67 GHz can be increased by decreasing the temperature of the receiver. It can be cryogenically cooled owing to cryogenic receiver technology already being in use as a key technique in space radio telescopes.

\item[6)] At frequency band $\geq$74 MHz, the telescope type can be either space truss-structure expandable antenna, or space umbrella-shape expandable antenna. At the lowest frequency band of 30 MHz, three orthogonal dipole antennas would be considered, instead of the dish-type telescope.  Either one requires large diameters to achieve high sensitivity. The opening of the folded space antennas in space is a key technique under investigation.
\end{enumerate}

\section{Scientific objectives}
\label{sec:science_objectives}

The low frequency radio Universe offers exciting prospects to gather a coherent view of physical phenomena and enable discoveries unique to this domain.  The low frequency space VLBI observations can benefit from high resolution, absence of ionospheric disturbances on the space-ground and space-space baselines, as well as long integration time. 

The achievable baseline sensitivity can be estimated from the telescope parameters (e.g., the system equivalent flux density, SEFD) and the observation setup (e.g., integration time, bandwidth). At 1.67 GHz, 1-$\sigma$ baseline sensitivity within 10 min integration time of CM1 is 0.09 mJy (with FAST), 0.23 mJy (with Effelsberg), and 0.065 mJy (with SKA1-Mid). At 330 MHz, the baseline sensitivity is 0.41 mJy (CM1-FAST), 2.9 mJy (CM1-Effelsberg), 0.36 mJy (CM1-SKA1-Mid), 0.24 mJy (CM1-SKA1-Low). Setting a 7-$\sigma$ threshold, the CM mission can easily detect 1 mJy compact source at 1.67 GHz and a few mJy source at 330 MHz. We should mention that practical sensitivity calculation should be modified (increased) by considering the coherency of the ground radio telescopes and the means of the time and frequency metrology used for the space telescopes, and the ionospheric calibration errors of ground-based telescopes. The relatively larger beam size at low frequency add the promising capability for in-beam phase referencing. Moreover the utility of MultiView calibration enables high-precision astrometric space VLBI \cite{ref19}, that requires multiple calibrators and simultaneous observations; thus the multi-beam capability of the space telescope is highly desirable, while the technical feasibility of the multi-beam device on the satellite remains a concern.  This unprecedented high sensitivity will undoubtedly open a new discover space in many science disciplines.

\begin{figure}[!t] 
\centering
\begin{subfigure}{.3\columnwidth}
  \centering
  \includegraphics[width=1.0\linewidth]{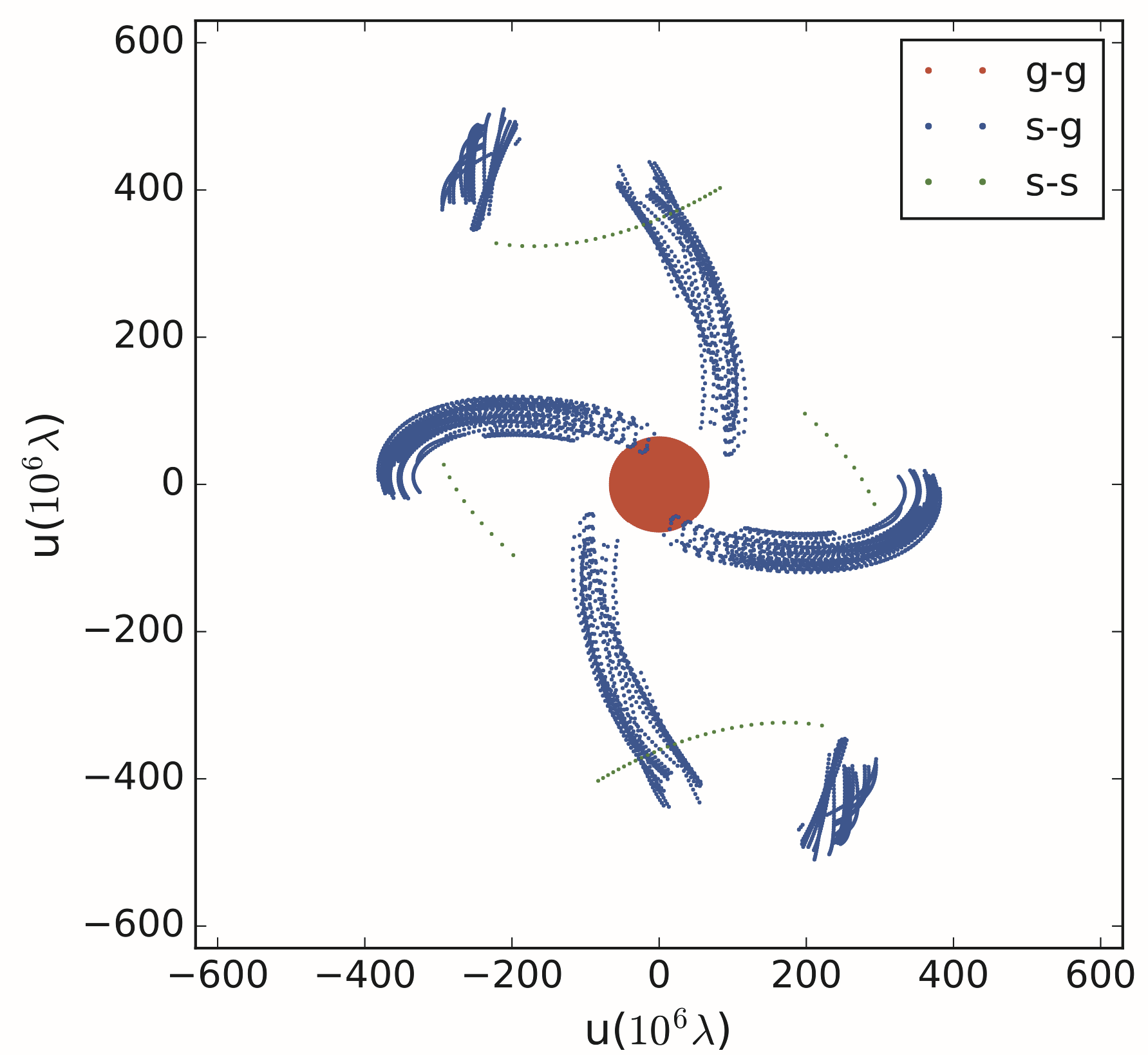}
\end{subfigure}%
\begin{subfigure}{.3\columnwidth}
  \centering
  \includegraphics[width=1.0\linewidth]{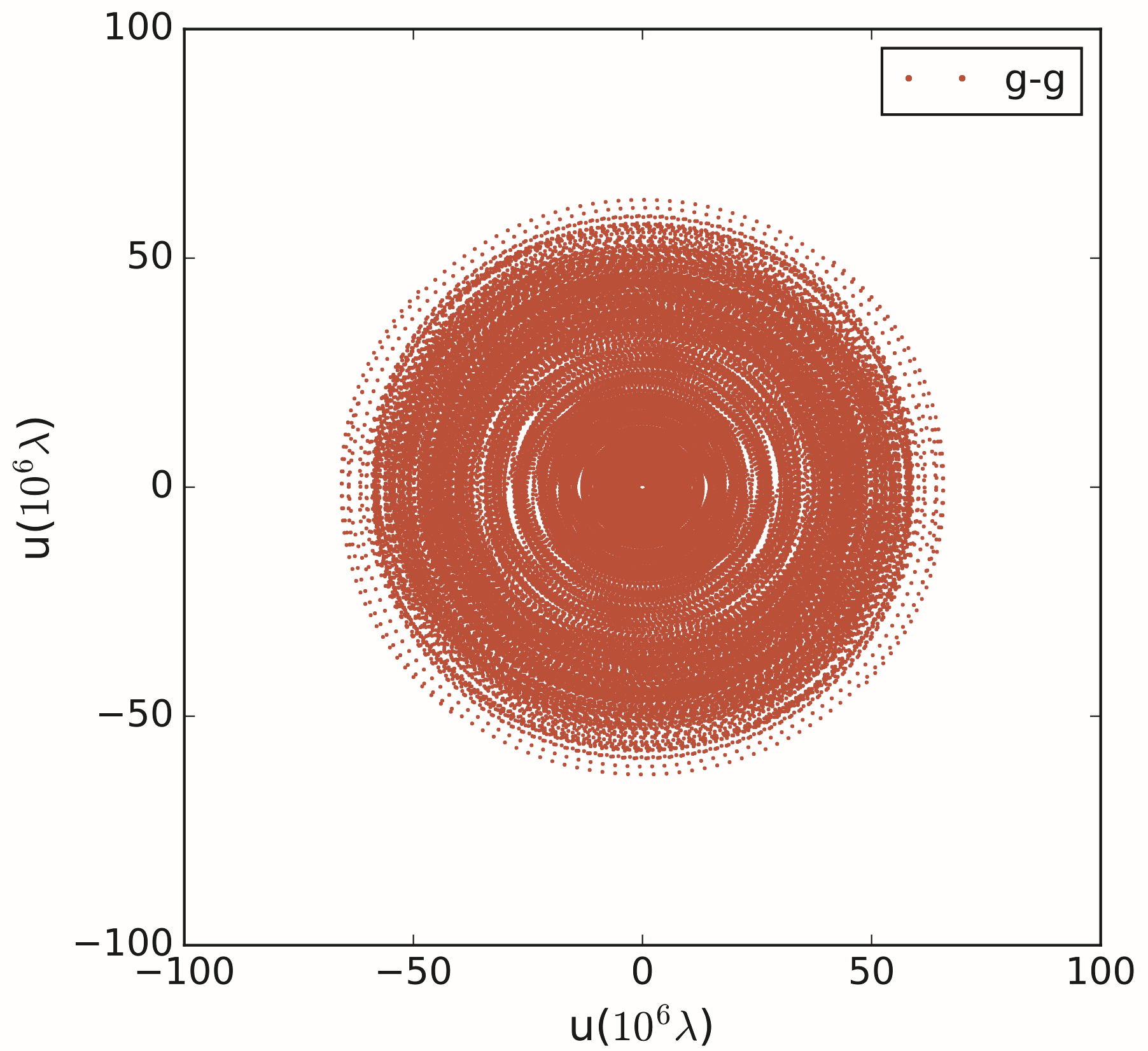}
\end{subfigure}
\begin{subfigure}{.3\columnwidth}
  \centering
  \includegraphics[width=1.0\linewidth]{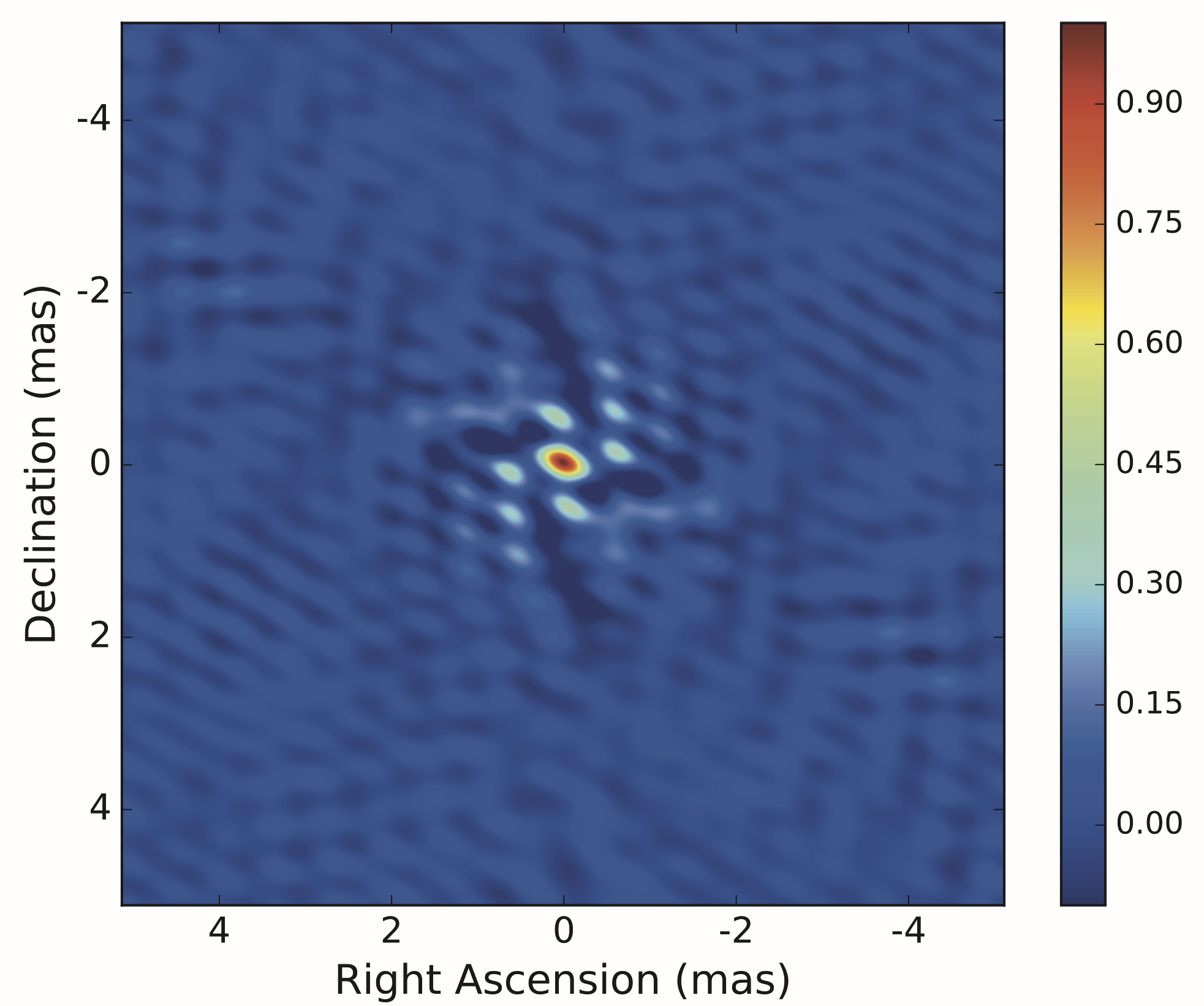}
\end{subfigure}
\caption{($u,v$) coverage of CM (left), ground telescopes only  (middle) and dirty beam (right).  The observing frequency used here is 1.67 GHz.  The target is a high-declination source, J0906+6930, a high-redshift quasar \citep{ref20}. The projected $u$ and $v$ distances are in unit of observing wavelength (i.e., 18 cm). Telescopes used in this simulation are: CM1, CM2, FAST, EVN, and VLBA. The observing time is assumed on 2023 January 1. The orbital parameters are referred to Table ~\ref{tab:tab1}. Green colored lines (s-s) show the space-space baselines, blue colored lines (s-g) show the space-ground baselines, and red colored lines (g-g) the ground-only baselines.  The right panel demonstrates the dirty beam, 0.45 mas $\times$ 0.28 mas, at a position angle of 65.5 degree. The ($u,v$) plot and dirty beam simulation are made by using the software tool UVSIM \cite{ref26} and VNSIM\cite{ref27}.\label{fig:fig2}}
\end{figure}

The ($u,v$) coverage represents the imaging performance of a VLBI network. Figure~\ref{fig:fig2} demonstrates the expected ($u,v$) coverage of the CM. As may be seen, the CM gains an advantage over previous space VLBI missions by an enriched coverage in intermediate ($u,v$) spacing,  important to unravel coherent source structures. Figure ~\ref{fig:fig2} shows the ($u,v$) coverage of the CM mission and the generated dirty map. Unlike the VSOP and RadioAstron which have a narrow elliptical beam, the CM benefits from two space telescopes, and forms an elliptical beam with axis ratio of ~1.7:1. The sidelobe of the dirty beam is greatly suppressed in this manner.  These improvements are useful for large snap-shot sky survey, and/or tracing fast variable radio sources which require suitable imaging quality within a short time period. An estimate based on a network consisting of CM1, CM2, FAST, EVN, and SKA1-Mid gives an image rms noise of 1.7 microJy/beam at 1.67 GHz with an hour of integration, an order of magnitude better than EVN only, besides the resolution improvement by a factor of $\sim$10. Inclusion of additional telescopes will further decrease the noise level. Such an image sensitivity, along with the mas-scale resolution, enables the detection of a large number of SMBHBs, fast radio bursts, afterglow from neutron-star mergers, and nuclear starbursts.

Key science drivers at and below 1.7 GHz VLBI broadly include:
\begin{enumerate}
\item Study of the supermassive black holes (SMBHs) across the cosmic epoch.  This includes the formation and growth of the first-generation SMBHs in the early Universe, their merger events and binary evolution, and the impact of AGN on the galaxy evolution. The current VLBI observations of high-redshift quasars are limited to a handful of bright sources. Other than J0906+6930\cite{ref20} (the brightest $z > 5$ radio quasar) and a few others, the remaining high-redshift quasars are typically faint ($<$1 mJy) and their radio structure is extremely compact, strongly necessitating the high-sensitivity high-resolution space VLBI as proposed here.  

\item 
Some AGN cores show brightness temperatures significantly higher than the inverse Compton catastrophe limit \cite{ref21}. Though the reason for that still remains an open question, any interpretation may invoke a change of the current understanding for the jet models. The highest brightness temperature is related to the maximum baseline length of the VLBI network, regardless of the observing frequency and resolution. Therefore the brightness temperature limit can only be measured on baselines beyond the Earth, {\it i.e.}, the space-earth or space-space baselines.  

\item The coalescence of SMBHBs contribute to the nanoHz gravitational wave (GW) background. Understanding their distribution and evolution will significantly contribute to GW studies.  The observational evidence of such a phenomenon for SMBHBs at the centers of galaxies (expected in a scenario involving major mergers) requires both high sensitivity and high angular resolution, which can only be achieved through space VLBI \cite{ref22}.

\item Search, monitoring and characterization of radio transients (including supernovae, gamma-ray bursts, pulsars, variable stars and tidal disruption events related to re-activated SMBHs) and synergies with multi-messenger astronomical studies by either contributing novel sources (serendipitous discoveries) or as aiding in follow-up observations of transient electromagnetic counterparts (such as that from GW170817 event \cite{ref23,ref24}). VLBI offers unprecedented high-accuracy position for these transient sources.

\item Discovering new nearby extra-solar Jupiter-like gas rich planets which could possess strong magnetic fields causing plasma interactions and low frequency radio emission associated with magnetospheric activity. Jupiter is observed to often show low frequency radio bursts; most of this is polarized implying an underlying magnetic field.  The driver of this radio emission could involve a complex interplay between the rotation of Jupiter, its intrinsic magnetic field and the solar wind/bursts with specific details not clearly understood. Space VLBI can sufficiently resolve surface features of Jupiter and thus vastly improve our understanding of this emission, also by observing a large sample of Jupiter-like exoplanets. Due to their compactness, high resolution (50 mas) is needed to resolve extra-solar and Jupiter-like planets at a distance of 100 pc, corresponding to a baseline of $>$40,000 km at 30 MHz \cite{ref25}. 

\item Studies of the solar wind and environment in our local solar neighborhood enable investigations of interplanetary scintillation and provide inputs for space weather and exploration.

\end{enumerate}

There are also other broad science applications.  
Imaging the cores of quasars and radio galaxies in nearby Universe can help elucidate the formation and collimation of their jets. The space VLBI capability can be leveraged to produce high resolution images of target sources and provide vastly better astrometric positions (e.g. pulsars and other transients) compared to terrestrial low frequency arrays. Observations of cosmic hydroxyl (OH) masers at 1.67 GHz can help to better understand the star formation in the Milky Way and trace the star formation history in the Universe up to redshifts of 2-3.  
One concern associated with low-frequency VLBI imaging is the scattering of the radio waves by the interstellar and interplanetary medium, that leads to enlarge the apparent size of a compact source. The broadening of the source image is proportional to $\lambda^2$, where $\lambda$ is the observation wavelength; thus this effect becomes stronger at lower frequencies. In addition, the degree of the scattering is a function of direction on the plane of the sky, becoming stronger towards the direction of the Galactic center and weaker at high Galactic latitudes. These factors should be taken into account in low-frequency VLBI observations.
On the other hand, the neutral hydrogen (HI) lines originating from the EoR are extended emission in nature, therefore they should be observed in the single-dish mode when the telescopes are far away from the Earth, eliminating the RFI contamination.  

\section{Summary }
\label{sec:summary }
The Shanghai Astronomical Observatory of the Chinese Academy of Sciences is proposing a space-based VLBI mission (Cosmic Microscope) for operation in the low frequency regime.  It is expected to provide a high angular resolution (smaller than 1 mas) and a gain in sensitivity unmatched by earlier or contemporary astronomical measurement technique. The technical requirements and science objectives for the proposed space VLBI mission are outlined. Combining this with ground largest telescopes (FAST, Arecibo, SKA1, GMRT, JVLA), the CM mission can achieve baseline sensitivity of a few mJy and imaging noise of microJy/beam level. Compared to earlier space VLBI programmes, the baseline sensitivity of the CM is improved by a factor of 10 or higher owing to the addition of newly constructed giant ground telescopes, e.g., FAST and SKA1, and the resolution is 10 times higher than ground-only VLBI networks.  These attributes are essential to unravel the fine features of compact celestial objects and to better understand associated physical phenomena. The combination of two space telescopes significantly improves the ($u,v$) coverage on space-ground baselines, and consequently the image performance. In the long run, three or more space telescopes are more desirable to perform imaging space-based telescopes only. CM expects to detect the weak and compact population of Galactic (e.g., GW counterparts, fast radio bursts, pulsars) and extragalactic radio sources (e.g., SMBHBs) which are not very well understood yet. The space VLBI at the 30--300MHz frequency range opens a new electromagnetic window to observe the radio emission from Jovian-type exoplanets, marking an important step in exoplanet studies. These pioneer scientific objectives can only be achieved through the space VLBI technique. 

\section*{Acknowledgments}
The authors thank Profs. Shuhua Ye, Zhihan Qian, Xiaoyu Hong, Weimin Zheng and other CM team members for their constructive comments, thank Ken Kellermann, Leonid Gurvits, Richard Dodson and Maria Rioja for useful discussion on the mission concept, and Zhiyu Cao, Wei Wang, Xiaofei Ma for their helpful discussion on the satellite techniques.


\bibliographystyle{elsarticle-num}



\end{document}